\documentclass[twocolumn]{jpsj3} 

\newcommand{\lsim}
 {\ \raise.35ex\hbox{$<$}\kern-0.75em\lower.5ex\hbox{$\sim$}\ }
\newcommand{\gsim}
 {\ \raise.35ex\hbox{$>$}\kern-0.75em\lower.5ex\hbox{$\sim$}\ }
\def\journal #1#2#3#4{#1 {\bf #2} (#4) #3}
\def\PR{Phys.\ Rev.}
\def\PRA{Phys.\ Rev.\ A}
\def\PRB{Phys.\ Rev.\ B}
\def\PRL{Phys.\ Rev.\ Lett.}

\def\PRS{Proc.\ Roy.\ Soc.}

\def\IJMP{Int.\ J.\ Mod.\ Phys.}

\def\JLTP{J.~Low Temp.~Phys.}

\def\JPSJ{J.\ Phys.\ Soc.\ Jpn.}

\def\NP{Nat.~Phys.}

\def\RMP{Rev.\ Mod.\ Phys.}
\def\PTP{Prog.\ Theor.\ Phys.}

\def\EPL{Europhys.\ Lett.}
\def\RPP{Rep.~Prog.~Phys.}
\title{Effects of Long-Range Correlations on Nonmagnetic Mott Transitions 
in Hubbard model on Square Lattice}

\author{Tomoaki {\sc Miyagawa}\thanks{E-mail address: 
                          miyagawa@cmpt.phys.tohoku.ac.jp} and 
        Hisatoshi {\sc Yokoyama} 
}
\inst{Department of Physics, Tohoku University, 
      Sendai 980-8578 
}
\abst{ 
The mechanism of Mott transition in the Hubbard model on the square 
lattice is studied without explicit introduction of magnetic and 
superconducting correlations, using a variational Monte Carlo method. 
In the trial wave functions, we consider various types of binding 
factors between a doubly-occupied site (doublon, D) and an empty 
site (holon, H), like a long-range type as well as a conventional 
nearest-neighbor type, and add independent long-range D-D (H-H) factors. 
It is found that a wide choice of D-H binding factor leads to 
Mott transitions at critical values near the band width. 
We renew the D-H binding picture of Mott transitions by introducing 
two characteristic length scales, the D-H binding length $\ell_{\rm DH}$ 
and the minimum D-D distance $\ell_{\rm DD}$, which we appropriately 
estimate. 
A Mott transition takes place at $\ell_{\rm DH}=\ell_{\rm DD}$. 
In the metallic regime ($\ell_{\rm DH}>\ell_{\rm DD}$), the domains 
of D-H pairs overlap with one another, thereby doublons and holons can 
move independently by exchanging the partners one after another. 
In contrast, the D-D factors give only a minor contribution 
to the Mott transition. 
}

\kword{Mott transition, Hubbard model, variational Monte Carlo, 
doublon-holon binding, square lattice}

\begin{document}
\maketitle

\section{Introduction\label{sec:intro}}
The Mott transitions\cite{Mott} free from the spin degree of freedom 
were recently realized in ultracold bosonic atoms on optical 
lattices\cite{Greiner,1D,Kohl,Spielman}. 
In the Bose Hubbard models,\cite{Fisher} which faithfully 
describe these systems,\cite{Jaksch,Bloch-Rev} superfluid-insulator 
(Mott) transitions occur when kinetic and interaction energies are 
competitive.\cite{critical,Uc-1d,QMC-2D1,QMC-2D2,QMC-2D3,Monien,Uc-3d1,Uc-3d2} 
It follows that the essence of ``Mott physics" can be separated 
from magnetic metal-insulator transitions,\cite{Slater} 
often arising in weakly correlated regimes in half-filled-band 
electronic systems, especially, with good nesting conditions. 
Aside from the spinless bosons, systems like organic superconductors 
$\kappa$-(BEDT-TTF) salts\cite{ET} and the cuprate superconductors 
as doped Mott insulators\cite{PALee,OF,YOTKT} require a deep 
understanding about the mechanism of nonmagnetic Mott transitions. 
Thus, it is significant to shed light on the Mott transition between 
virtual paramagnetic phases in the Hubbard model on the square 
lattice, which, though, actually has an antiferromagnetic (AF) long-range 
order for any finite value of positive $U/t$ ($U$: onsite correlation 
strength; $t$: hopping integral).\cite{Hirsch,YS2} 
\par 

The variation theory\cite{Gutz,BR} has long been one of the main streams 
to study ground-state properties of the Mott transition. 
In particular, the variational Monte Carlo (VMC) 
approaches\cite{McMillan,Ceperley,YS1} are effective for its reliability 
in dealing with the local correlation and wide applicability. 
In the previous VMC studies related to paramagnetic Mott transitions, 
the following properties have been clarified. 
(i) The well-known Gutzwiller wave function,\cite{Gutz} with only 
onsite correlation, does not undergo a Mott transition for finite $U/t$ 
and in finite dimensions. 
Namely, the Brinkman-Rice transition\cite{GA,BR} is unreal.\cite{YS1,MV} 
(ii) Wave functions with short-range intersite attractive correlations 
between a doubly-occupied site (doublon, D) and an empty site 
(holon, H)\cite{Kaplan,Fazekas,YS},
which are minus and plus carriers in the neutral background (singly occupied sites), 
can properly describe the Mott transition.\cite{Y-PTP,YTOT,Watanabe,YOT} 
In particular in ref.~\citen{YOT}, the D-H binding-unbinding mechanism 
is studied for a projected $d$-wave singlet state as well as a projected 
Fermi sea for the two-dimensional Hubbard model ($t$-$t'$-$U$ model).
A similar result was also reached using a Jastrow-type wave 
function.\cite{Capello} 
Later, we have found in two dimensions that Mott transitions occur 
even in the wave functions in which a doublon must be necessarily 
accompanied by at least one holon in the nearest-neighbor (NN) sites. 
Namely, the state in which a doublon and a holon always tightly bind 
one another can be metallic (see \S\ref{sec:mecbind}). 
This finding requires a modification of the above picture\cite{YOT} 
of Mott transitions through D-H binding and a simple release from it. 
\par

In this paper, we address the following subjects by applying a VMC 
method to the half-filled-band Hubbard model on the square lattice: 
(i) We extend the D-H attractive correlation factors so as to include 
some long-range types, and corroborate the decisive effect of D-H binding 
on nonmagnetic metal-insulator transitions. 
(ii) We introduce long-range D-D (and H-H) factors independently 
of the above D-H factors. 
Thereby, we can distinguish the roles of the two factors for the Mott 
transition. 
(iii) We generalize the picture of the Mott transition so as to comprehend 
the one arising in the completely D-H bound state, and to treat it 
more quantitatively. 
By checking various wave functions, we are convinced that this conception 
is applicable to a wide range of systems, including Bose Hubbard 
models.\cite{YMO} 
\par

This paper is organized as follows: 
In \S\ref{sec:formu}, the method used in this paper is formulated.  
In \S\ref{sec:result}, we discuss various aspects of the VMC results.
In \S\ref{sec:mecha}, we propose an improved picture of Mott transitions, 
and confirm its applicability. 
Section \ref{sec:summa} is assigned to summary.
\par 

A part of the results in this study was published before.\cite{Miyagawa}

\section{Formulation\label{sec:formu}}
After brief introduction of the Hubbard model in \S\ref{sec:model}, 
in \S\ref{sec:vmf}, we describe the trial wave functions treated 
in this paper. 
In \S\ref{sec:calccond}, we outline the VMC calculations.
\par

\subsection{Hubbard model on square lattice\label{sec:model}}
We study the single-band Hubbard model with 
$t,U\geqq0$,\cite{Gutz,Hubbard,Kanamori} which is fundamental 
to describe the physics of Mott transitions:\cite{DMFT} 
\begin{eqnarray}
H &=& H_t + H_U \nonumber \\
  &=& -t \sum_{\langle i,j\rangle ,\sigma} {(c_{i\sigma}^\dag c_{j\sigma} 
+ c_{j\sigma}^\dag c_{i\sigma})} 
+ U\sum_i d_i, \qquad
\label{eq:model} 
\end{eqnarray} 
where $c_{i\sigma}$ is an electron annihilation operator of site 
$i$ and spin $\sigma$, $d_i=n_{i\uparrow}n_{i\downarrow}$ and 
$n_{i\sigma}=c_{i\sigma}^\dag c_{i\sigma}$. 
In this paper, we focus on the case of half filling on the square lattice 
with only the nearest neighbor (NN) hopping, 
\begin{eqnarray}
H_t &=& \sum_k \epsilon_k c_{k\sigma}^\dag c_{k\sigma}, 
\label{eq:H_t} \\
\epsilon_k &=& -2t ( \cos k_x + \cos k_y ),
\label{eq:epsilon} 
\end{eqnarray} 
and use $t$ as the energy unit. 
In applying a variational Monte Carlo (VMC) method to this model, 
we use finite-size systems of $L\times L$ ($=N_{\rm s}$) sites 
up to $L=18$ ($N_{\rm s}=324$) with the periodic and antiperiodic 
boundary conditions in $x$ and $y$ directions, respectively, 
to meet the closed-shell condition.
\par

\subsection{Variational wave functions\label{sec:vmf}}
In \S\ref{sec:A(NN)}, we explain the conventional NN D-H binding 
wave function and the completely D-H bound state as its limiting case. 
In \S\ref{sec:JSF}, we introduce a series of wave functions with 
long-range D-H binding factors and independent D-D (H-H) correlation 
factors. 
\par

\subsubsection{Nearest-neighbor correlation factor\label{sec:A(NN)}}
%
The simplest but fundamental trial wave function is the celebrated 
Gutzwiller wave function (GWF),\cite{Gutz}
\begin{equation}
\Psi_{\rm G} = P_{\rm G} \Phi_{\rm F},
\label{eq:GWF} 
\end{equation}
where $\Phi_{\rm F}$ is the Fermi sea, and 
\begin{equation}
P_{\rm G} = \prod_j \left[ 1 - ( 1 - g )  d_j \right]. 
\label{eq:PGWF} 
\end{equation}
Here, $g$ is the variational parameter which adjusts the doublon 
density, $d=\sum_j\langle d_j\rangle/N_{\rm s}$. 
It is known that GWF is metallic for $U/t<\infty$.\cite{YS1} 
To describe the Mott transition, a wave function\cite{Kaplan,YS,Y-PTP} 
with a NN D-H binding correlation\cite{Castellani} has been often used:
\begin{equation}
\Psi_{\rm A(NN)} = P_{\rm A}^{\rm NN}\Psi_{\rm G} 
= \prod_j \left( 1 - \mu \hat{Q}_j \right)\Psi_{\rm G},
\label{eq:PA(NN)}
\end{equation}
\begin{equation}
\hat{Q}_j =  
d_j \prod_{\vec{\tau}}( 1 - h_{j+\vec{\tau}} ) +
h_j \prod_{\vec{\tau}}( 1 - d_{j+\vec{\tau}} ),
\label{eq:Q(NN)}
\end{equation}
where $h_j= (1-n_{j\uparrow})(1-n_{j\downarrow})$, 
$\mu$ ($0\le\mu\le 1$) is a variational parameter 
which controls the number of isolated doublons (D without H in its 
NN sites) and isolated holons, and $\vec{\tau}$ runs over the four 
NN sites of the site $j$. 
For $\mu=0$, $\Psi_{\rm A(NN)}$ is reduced to $\Psi_{\rm G}$. 
In the other limit, $\mu=1$, $\Psi_{\rm A(NN)}$ becomes the completely 
D-H bound state,
\begin{equation}
\Psi_{\rm A(bind)} = P_{\rm A}^{\rm bind}\Phi_{\rm G}
= \prod_j \left\{ 1 - \hat{Q}_j \right\}\Phi_{\rm G}. 
\label{eq:FA(bind)}
\end{equation}
In $\Psi_{\rm A(bind)}$, a doublon (holon) must be accompanied by 
at least one holon (doublon) in its NN sites, so that, superficially, 
$\Psi_{\rm A(bind)}$ always seems insulating. 
However, it turns out to be metallic for small $U/t$, because a doublon 
(holon) can have multiple holons (doublons) in its NN sites, as will be 
discussed in \S\ref{sec:mecbind}. 
\par

\subsubsection{Long-range Jastrow factors\label{sec:JSF}}
According to the result of exact diagonalization in the one-dimensional 
half-filled-band Hubbard model,\cite{YS} the ground state of which is 
a paramagnetic insulator for any positive $U/t$,\cite{Lieb-Wu} 
the magnitude of the coefficient of the basis having only one D-H pair 
with the D-H distance $r$ decreases exponentially with $r$ for large $U/t$. 
Assuming a similar situation arises in two dimensions, we introduce 
several types of long-range D-H attractive (A) correlation factors 
$P_{\rm A}$, and the corresponding D-D and H-H repulsive (R) factors 
$P_{\rm R}$: 
\begin{eqnarray}
\nonumber
P_{\rm A} = \prod_j \left( f_{\rm A}\left(|\vec{r}_j^{\rm A}|\right) 
\left\{ 
d_j \left[ 1-\prod_{\vec{r}\in\{{\vec{r}_j^{\rm A}}\}}\left( 1 - h_{j+\vec{r}} \right) \right]
\right. \right.  \\ +  \left. \left.
h_j \left[ 1-\prod_{\vec{r}\in\{\vec{r}_j^{\rm A}\}}\left( 1 - d_{j+\vec{r}} \right) \right]
\right\} \right), 
\label{eq:PA(jas)}  
\end{eqnarray}
\begin{eqnarray}
\nonumber
P_{\rm R} = \prod_j \left( f_{\rm R}\left(|\vec{r}_j^{\rm R}|\right) 
\left\{ 
d_j \left[ 1-\prod_{\vec{r}\in\{{\vec{r}_j^{\rm R}}\}}\left( 1 - d_{j+\vec{r}} \right) \right]
\right. \right.  \\ +  \left. \left.
h_j \left[ 1-\prod_{\vec{r}\in\{{\vec{r}_j^{\rm R}}\}}\left( 1 - h_{j+\vec{r}} \right) \right]
\right\} \right), 
\label{eq:PR(jas)}
\end{eqnarray}
where the index $j$ of the outer product runs over all the sites, and 
$\vec{r}_j^{\rm A}$ ($\vec{r}_j^{\rm R}$) indicates the vector 
from the site $j$ to the nearest partner. 
Namely, we disregard more distant ones, so that the index $\vec{r}$ 
of the products in eq.~(\ref{eq:PA(jas)}) runs over only the sites 
of the nearest D-to-H (in the first term) and H-to-D (in the second term) 
distances. 
This choice is reasonable in view of the strong-coupling 
expansion.\cite{Harris}
Correspondingly, we treat the product of $\vec{r}$ in eq.~(\ref{eq:PR(jas)}) 
in a similar way. 
To measure distances, we adopt the stepwise or Manhattan metric, 
in which, for instance, $r=2$ for $(i,j)\leftrightarrow(i+1,j+1)$, and 
the range is $1\le r\le L$. 
For each correlation, we consider the three forms: 
\begin{equation}
f_{\rm A}(r) = \left\{
\begin{array}{ll} 
\displaystyle
\exp \left( - \frac{r-1}{\xi}\right),& \mbox{(a) exponential}\\
\displaystyle
\frac{1}{r^{\xi}},                   & \mbox{(b) power law}\\
\displaystyle
\xi_r,                               & \mbox{(c) optimizing}\\
\end{array}
\right. 
\label{eq:f(A)}  
\end{equation}
\begin{equation}
f_{\rm R}(r) = \left\{
\begin{array}{ll}
\displaystyle
1-\alpha \exp\left(- \frac{r-1}{\beta} \right), & \mbox{(a) exponential}\\
\displaystyle
1-\frac{\alpha}{r^{\beta}},                     & \mbox{(b) power law}\\
\displaystyle
\alpha_r,                                       & \mbox{(c) optimizing}\\
\end{array}
\right. 
\label{eq:f(R)}  
\end{equation}
where we fix $f_{\rm A}(1)$ and $f_{\rm R}(\infty)$ at unity 
for (a) and (b).
\par 

\vspace{-0.2cm}
\begin{figure}[hob]
\begin{center}
\includegraphics[width=80mm,clip]{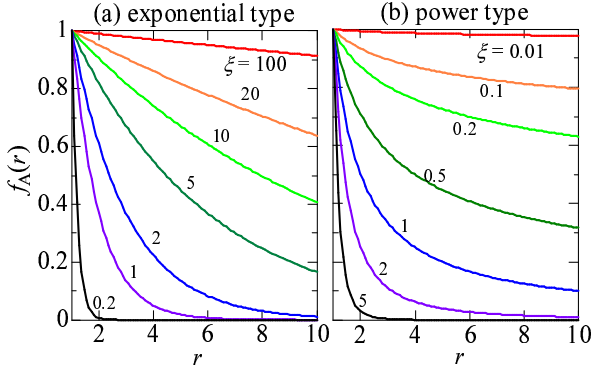}
\end{center}
\vskip -5mm
\caption{
(Color online) Weight of long-range D-H attractive correlations
as a function of distance between D and H; 
(a) an exponentially decaying type [eq.~\ref{eq:f(A)}(a)], and 
(b) a power-law decaying type [eq.~\ref{eq:f(A)}(b)]. 
}
\label{fig:01fdh}
\end{figure}
%

In eq.~(\ref{eq:f(A)}), each $\xi$ ($\xi_r$) is a variational parameter. 
We consider two typical forms: (a) exponentially decaying and 
(b) power-law decaying, as shown in Fig.~\ref{fig:01fdh}. 
Furthermore, in (c), we do not assume a specific form of $f_{\rm A}(r)$, 
and optimize all $\xi_r$'s ($2\le r\le L$) simultaneously as variational 
parameters. 
The type (c) is the best form of $f_{\rm A}(r)$, here; the other types 
are specific cases of (c). 
In eq.~(\ref{eq:f(R)}), we again assume the repulsive correlation 
becomes weaker, as $r$ increases. 
Corresponding to $f_{\rm A}(r)$, we consider three forms for $f_{\rm R}(r)$: 
(a) an exponentially and (b) a power-law decaying types, both
with the parameters $\alpha$ adjusting the weight of $r=1$, 
and $\beta$ controlling the decaying length. 
In (c), we optimize all $\alpha_r$'s ($1\le r\le L$) simultaneously. 
Notice that here we permit $\alpha_r$ to be larger than 1, meaning 
$P_{\rm R}$ possibly works as D-D (H-H) {\it attractive} correlations. 
\par

\begin{table}[hob]
\caption{
Summary of trial wave functions. 
In the second column, we abbreviate the type of correlation, with 
GW being the onsite (Gutzwiller) repulsion. 
}
\begin{center}
\begin{tabular}{c|c|c|c}
\hline
$\Psi$       & correlation & correlation & param. \\
abbreviation &     type    & range       & number \\
\hline\hline
 GWF     & GW    & onsite & 1  \\
\hline
 A(NN)   & GW+DH & NN  & 2  \\
 A(bind) & GW+DH & NN  & 1  \\
\hline
 A(exp)  & GW+DH    & long & 2   \\
 R(exp)  & GW+DD    & long & 3   \\
 AR(exp) & GW+DH+DD & long & 4   \\
\hline
 A(pow)  & GW+DH    & long & 2   \\
 R(pow)  & GW+DD    & long & 3   \\
 AR(pow) & GW+DH+DD & long & 4   \\
\hline
 A(opt)  & GW+DH    & long & $L$   \\
 R(opt)  & GW+DD    & long & $L$+1 \\
 AR(opt) & GW+DH+DD & long & 2$L$  \\
\hline
\end{tabular}
\end{center}
\label{table:index}
\end{table}
%
In this work, we study a series of wave functions by combining 
the above correlation factors as follows:
\begin{equation}
\begin{array}{l}
\;\;\Psi_{\rm A(exp)}  = P_{\rm A}^{\rm(a)}\Psi_{\rm G}, \\
\;\;\Psi_{\rm R(exp)}  = P_{\rm R}^{\rm(a)}\Psi_{\rm G}, \\
\Psi_{\rm AR(exp)} = P_{\rm A}^{\rm(a)}P_{\rm R}^{\rm(a)}\Psi_{\rm G}, 
\end{array}
\label{eq:Psiexp}  
\end{equation}
\begin{equation}
\begin{array}{l}
\;\;\Psi_{\rm A(pow)}  = P_{\rm A}^{\rm(b)}\Psi_{\rm G}, \\
\;\;\Psi_{\rm R(pow)}  = P_{\rm R}^{\rm(b)}\Psi_{\rm G}, \\
\Psi_{\rm AR(pow)} = P_{\rm A}^{\rm(b)}P_{\rm R}^{\rm(b)}\Psi_{\rm G}, 
\end{array}
\label{eq:Psipow}
\end{equation}
\begin{equation}
\begin{array}{l}
\;\;\Psi_{\rm A(opt)}  = P_{\rm A}^{\rm(c)}\Psi_{\rm G}, \\
\;\;\Psi_{\rm R(opt)}  = P_{\rm R}^{\rm(c)}\Psi_{\rm G}, \\
\Psi_{\rm AR(opt)} = P_{\rm A}^{\rm(c)}P_{\rm R}^{\rm(c)}\Psi_{\rm G}, 
\end{array}
\label{eq:Psiopt}  
\end{equation}
where the superscripts of $P_{\rm A}$ and $P_{\rm R}$ correspond to 
the function types (a)-(c) in eqs.~(\ref{eq:f(A)}) and (\ref{eq:f(R)}). 
For $\Psi_{\rm AR}$, we unify the function types of $f_{\rm A}(r)$ and 
$f_{\rm R}(r)$, for simplicity. 
In Table.~\ref{table:index}, we summarize the used wave functions. 
\par

Note that the projector $P_{\rm A}^{\rm (c)}P_{\rm R}^{\rm (c)}$ 
in eq.~(\ref{eq:Psiopt}) is different 
from the Jastrow factor used in related papers,\cite{Capello-1D,Capello} 
especially in that the magnitude of long-range D-D factor in them is 
connected to the inverse of corresponding D-H attractive factor, so that 
the D-D factor is necessarily repulsive as far as the D-H factor 
is attractive. 
\par

\subsection{Variational Monte Carlo calculations\label{sec:calccond}}
In this subsection, we briefly describe the outline of VMC calculations 
implemented in this study.
\par

A correlated measurement or optimization-VMC technique\cite{Umrigar} 
is used to optimize variational parameters up to $2L$. 
In the non-linear minimization process of energy expectation values 
for the wave functions with many parameters, we adopt a quasi-Newton 
method, in which gradient vectors are effectively calculated by recently 
proposed formulae,\cite{Umrigar-Filippi} and Hessian matrices are 
approximated by Broyden-Flecher-Goldfarb-Shanno formula,\cite{Fletcher} 
the use of which does not affect the exactness of optimization itself. 
In coding, we refer to an algorithm offered by Ibaraki and 
Fukushima.\cite{Ibaraki} 
For wave functions with a few parameters, we use a simple linear 
optimization together.
\par

In both algorithms, parameters as well as energy converge typically 
after first several rounds of iteration with different fixed sample sets; 
in each set we generate typically $2.5\times 10^5$ particle configurations 
using Metropolis algorithm. 
After the convergence, we continue excess rounds (10-20 times) 
of iteration in the optimization process with successively renewed 
configuration sets.
We determine the optimized values by averaging the data obtained 
in the excess rounds; in averaging, we exclude scattered data 
beyond the range of twice the standard deviation. 
The optimal value is an average of substantially more than several 
million samples. 
The variational energy and significant parameters [$g$, $\mu$, $\xi$, 
etc.] are obtained with sufficient accuracy, but the determination 
of insignificant parameters [$f_{\rm A}(r)$ and $f_{\rm R}(r)$ with 
$r>8$, etc.] is difficult, because $E$ depends on them only very slightly, 
in other words, particle configurations determining them appear 
extremely rarely. 
Anyway, such parameters have little influence on $E$ and other quantities. 
Physical quantities are calculated typically with $2.5\times 10^5$ 
renewed configurations generated by the optimized parameter sets.
\par

Since in Mott critical regimes, the global minimum becomes more 
competitive with other local minima as $L$ increases, accurate energy 
minimization sometimes becomes not easy. 
This leads to the scattered data points near $U_c/t$ in some figures. 
\par

\section{Results\label{sec:result}}
In this section, we discuss the results of VMC calculations. 
In \S\ref{sec:energy} and \S\ref{sec:quantities}, energies and other 
quantities obtained by various wave functions are studied, respectively, 
in view of the Mott transition. 
In \S\ref{sec:MTSD}, the system-size dependence of the Mott critical 
values are discussed. 
In \S\ref{sec:D-H}, the differences among various D-H factors 
are studied. 
In \S\ref{sec:repulsive}, we discuss the effects of repulsive 
long-range factors. 
\par

\subsection{Behavior of energy}\label{sec:energy}
\begin{figure}[hob]
\begin{center}
\includegraphics[width=80mm,clip]{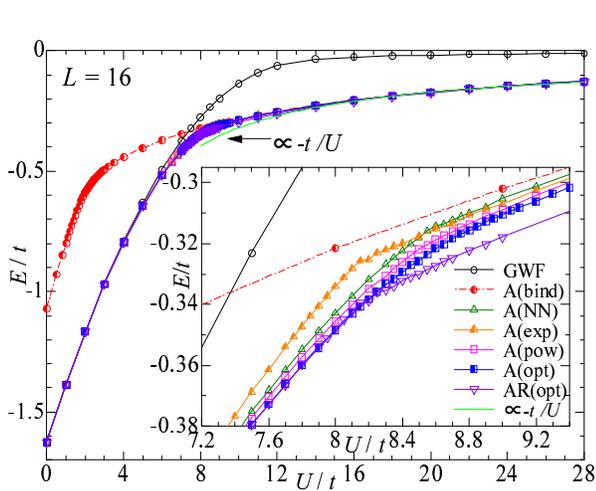}
\end{center}
\vskip -3mm
\caption{(Color online) 
Comparison of variational energies among various trial wave functions 
as function of correlation strength for $L=16$. 
The pale solid line is a guide line proportional to $-t/U$ expected 
from the strong-coupling expansion. 
The inset shows a magnification near the crossing point of GWF and 
A(bind).
}
\label{fig:02EU_Uc} 
\end{figure}
%
First of all, let us discuss the behavior of total energy per site 
$E/t$. 
In the main panel of Fig.~\ref{fig:02EU_Uc}, we compare $E/t$ 
among various wave functions in a wide range of $U/t$. 
Here, we notice GWF and the completely bound state $\Psi_{\rm A(bind)}$. 
It is known that GWF is metallic for $U/t<\infty$ and a relatively 
good for $U$ sufficiently smaller than the critical value 
of the Brinkman-Rice transition, $U_{\rm BR}=12.97t$.\cite{BR}
On the other hand, $\Psi_{\rm A(bind)}$ is insulating for $U/t\gsim 3$ 
(see \S\ref{sec:mecbind}), which fact is supported by the coincidence 
with the result of strong coupling expansion 
$\propto -t/U$,\cite{Harris} as shown in Fig.~\ref{fig:02EU_Uc}. 
$E/t$ of GWF is much lower than that of $\Psi_{\rm A(bind)}$ for $U/t<7.4$, 
while the relation is reversed for $U/t>7.4$, suggesting the phase 
switches from metal to insulator at $U/t\sim 7.4$. 
Now, we look at the D-H binding wave functions ($\Psi_{\rm A}$). 
$E/t$ of all $\Psi_{\rm A}$'s [$E({\rm A})$] except for 
$\Psi_{\rm A(bind)}$ behave similarly, 
namely, $E({\rm A})$ is very close to $E({\rm GWF})$ for $U/t\lsim 5$, 
somewhat smaller than both $E({\rm GWF})$ and $E(\Psi_{\rm A(bind)})$ 
for the intermediate values of $U/t$, and approaches 
$E(\Psi_{\rm A(bind)})$ for $U/t\gsim 9$. 
Hence, a wide class of D-H binding wave functions probably induces 
Mott transitions at $U\sim W$ [$W(=8t)$: band width]. 
This is consistent with the previous result for a short-range D-H 
binding wave function, in which a first-order Mott transition occur 
at $U_{\rm c}/t$=8.59 and 8.73 for $L=16$ and 18, respectively.\cite{YOT} 
\par

The detailed behavior of $E({\rm A})$ for intermediate $U/t$ is 
different among different types of D-H factors, as shown in the inset 
of Fig.~\ref{fig:02EU_Uc}. 
For example, $E(\Psi_{\rm A(exp)})$ exhibits a clear cusp at
$U\sim 8.12t$ $(\equiv U_{\rm c})$, and has a higher value than those 
of other $E({\rm A})$'s for $U<U_{\rm c}$, whereas $E(\Psi_{\rm A(opt)})$ 
exhibits smooth behavior in this regime, and has a lower value. 
The energy of the best function $\Psi_{\rm AR(opt)}$ is broadly 
similar to $E(\Psi_{\rm A(opt)})$ for $U\le U_{\rm c}$, but has 
an appreciably lower energy for $U>U_{\rm c}$. 
We will return to this subject in \S\ref{sec:D-H}. 
\par

\begin{figure}[!hob]
\begin{center}
\includegraphics[width=80mm,clip]{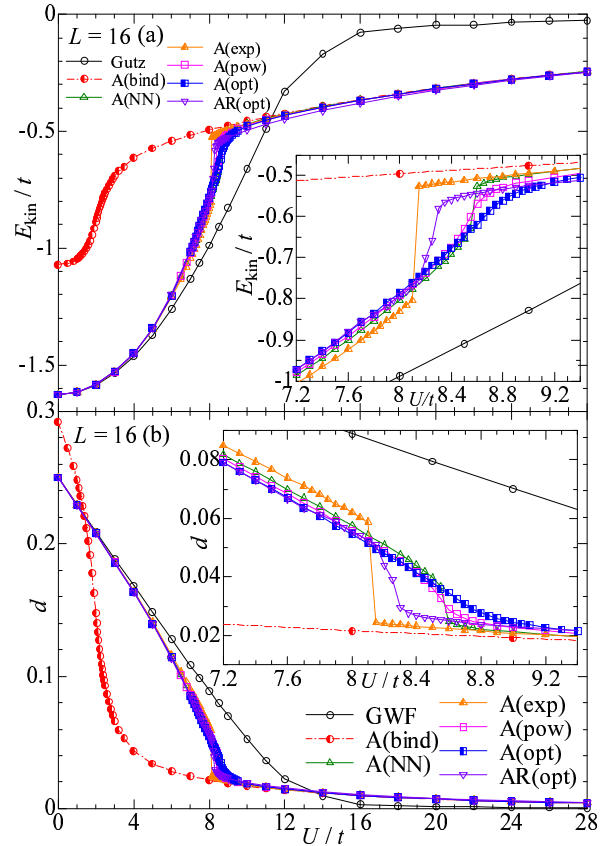}
\end{center}
\vskip -3mm
\caption{(Color online) 
The expectation values of (a) kinetic energy and (b) doublon density 
(interaction energy) are compared among the identical wave functions 
treated in Fig.~\ref{fig:02EU_Uc}. 
The insets show magnifications near the Mott critical points.
}
\label{fig:02EkinD}
\end{figure}
%
Next, we consider the components of energy. 
Although the total energy should be a continuous function of $U/t$, 
its components can exhibit stronger critical anomalies at $U_{\rm c}/t$ 
if the transition is first order. 
Figures \ref{fig:02EkinD}(a) and \ref{fig:02EkinD}(b) shows 
the kinetic energy, $E_{\rm kin}=\langle H_t\rangle/N_{\rm s}$, 
and doublon density (substantial interaction energy), 
$d=E_{\rm int}/U=\langle H_U\rangle/(UN_{\rm s})$, respectively. 
As evidently seen in the insets, $\Psi_{\rm A(exp)}$, which exhibits 
a clear cusp in $E/t$, has discontinuities at $U_{\rm c}/t$ in 
$E_{\rm kin}$ and $d$. 
This is an obvious sign of a first-order transition; actually we 
have observed a hysteresis around the critical point. 
Other wave functions behave more mildly at this system size, but 
the behavior evolves into more first-order-like as $L$ increases, 
as we will discuss in \S\ref{sec:MTSD}. 
The fact that $E_{\rm kin}$ ($d$) of every D-H binding wave function 
abruptly increases (decreases) in the critical region 
as $U/t$ increases is consistent with a requisite of the Mott 
transition: The metal-to-insulator transition is driven by reducing 
the interaction energy at the cost of the kinetic energy. 
This is in sharp contrast with antiferromagnetic and superconducting 
transitions arising in strongly correlated regimes.\cite{YTOT,YOTKT}
\par 

\subsection{Critical behavior of physical quantities\label{sec:quantities}}
To corroborate the realization of Mott transition in $\Psi_{\rm A}$
and estimate the critical value more accurately, we consider 
other physical quantities. 
\par

\begin{figure}[hob]
\begin{center}
\includegraphics[width=75mm,clip]{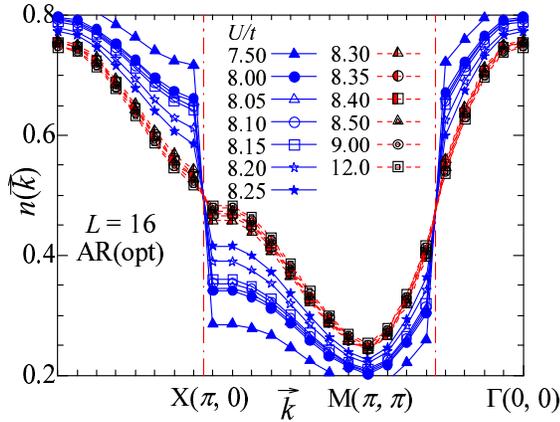}
\end{center}
\vskip -3mm 
\caption{(Color online)
The momentum distribution function is shown along the path, 
$(0,0)$$\rightarrow$$(\pi,0)$$\rightarrow$$(\pi,\pi)$$\rightarrow$$(0,0)$, 
calculated with the wave funcition AR(opt) for various values 
of $U/t$ near $U_{\rm c}/t\sim 8.27$ for $L=16$. 
The vertical dash-dotted lines indicate the positions of $\vec k_{\rm F}$ 
in the metallic cases. 
}
\label{fig:03nk}
\end{figure}
%
\begin{figure}[hob]
\begin{center}
\includegraphics[width=80mm,clip]{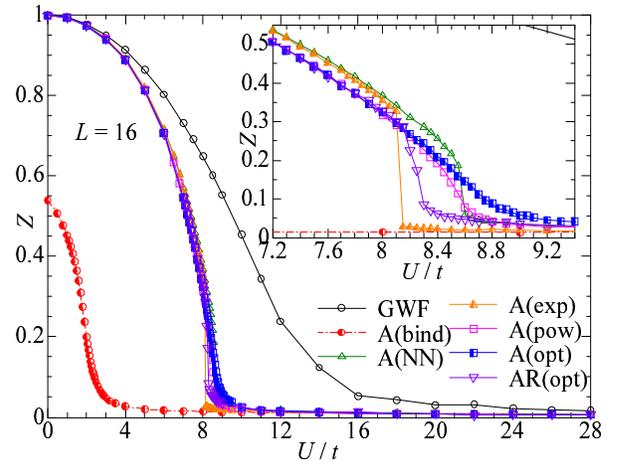}
\end{center}
\vskip -3mm
\caption{(Color online) 
Quasiparticle renormalization factor as function of $U/t$, estimated 
from discontinuities in $n(\vec k)$ at $\vec k=(\pi,0)$ for various type 
of wave functions. 
The inset shows the magnification near the Mott critical points.
A(bind) will be discussed later. 
}
\label{fig:03ZU_Uc}
\end{figure}
%
First, we discuss the momentum distribution function, 
\begin{equation}
n(\vec k) = \frac{1}{2} 
 \sum_{\sigma}\langle c_{\vec k\sigma}^\dag c_{\vec k\sigma} \rangle
     = \frac{1}{2N_s} \sum_{\vec j,\vec \ell,\sigma} e^{i\vec k\cdot\vec \ell} 
\left< c_{\vec j+\vec \ell\sigma}^\dag c_{\vec j\sigma} \right>. 
\label{eq:nk}
\end{equation}
%
In Fig.~\ref{fig:03nk}, we show, as a typical case, the result of 
$\Psi_{\rm AR(opt)}$, which exhibits first-order-like critical behavior 
in energy at $U_{\rm c}/t\sim 8.30$ (Fig.~\ref{fig:02EkinD}). 
Let us pay attention to the behavior around the Fermi surface 
near $\vec k=(\pi,0)$. 
For small values of $U/t$ ($<8.3$), a discontinuity of $n(\vec k)$ 
is evident at $\vec k_{\rm F}$, whereas the magnitude of 
discontinuity abruptly becomes small at $U\sim U_{\rm c}$ and 
remains very small for large $U/t$ ($>8.3$). 
To discuss quantitatively, we actually measure the jump of $n(\vec k)$, 
namely quasiparticle renormalization factor $Z$, at the X point 
$(\pi,0)$:
\begin{equation}
Z = n^{-}(k) |_{k \rightarrow X-0} 
  - n^{+}(k) |_{k \rightarrow X+0}, 
\label{eq:Z}
\end{equation}
where $n^{-}(k)$ [$n^{+}(k)$] is the fitting function of the segment 
$\Gamma$-X [X-M] of $n(\vec k)$ given by the third-order of least 
squares method. 
According to the Fermi liquid theory, $Z$ becomes zero for insulating 
states. 
As shown in Fig.~\ref{fig:03ZU_Uc}, $Z$ of every D-H binding function 
almost vanishes at $U/t=8.2$-9.0, although the detailed behavior somewhat 
differs among different wave functions. 
The small residual values for large $U/t$ are owing to the finite 
system sizes. 
Thus, we may regard the states for $U/t\gsim 9$ as insulating. 
As for $\Psi_{\rm AR(opt)}$, $Z$ most markedly drops at 8.2-8.3, 
which coincides with $U_{\rm c}/t$ evaluated from energy. 
In contrast, $Z$ of a metallic state GWF asymptotically approaches zero, 
as known.\cite{Y-PTP} 
\par

\begin{figure}[hob]
 \begin{center}
   \includegraphics[width=70mm,clip]{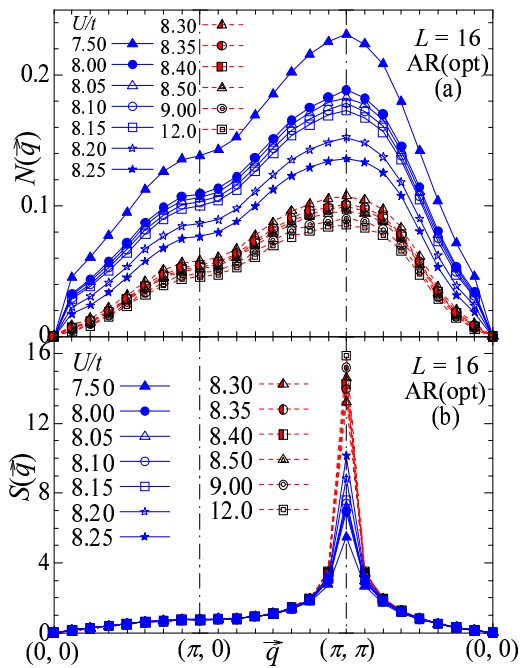}
 \end{center}
 \vskip -3mm
 \caption{(Color online)
(a) Charge density structure factor of AR(opt) for some values 
of $U/t$ near $U_{\rm c}/t\sim 8.3$. 
(b) Spin structure factor of AR(opt) for the same condition 
as (a). 
}
 \label{fig:03CQKSQK}
\end{figure}
%
Second, we consider the charge density correlation function 
in the wave-number space, 
\begin{equation}
N({\vec q})=\frac{1}{N_{\rm s}} 
\sum_{i,j}e^{i{\vec q}\cdot({\vec r}_i-{\vec r}_j)} 
\left\langle{n_{i} n_{j}}\right\rangle - n^2, 
\label{eq:nq}
\end{equation} 
which is known, within the variation theory, to behave as 
$N(\vec q)\propto |\vec q|$ for $|\vec q|\rightarrow 0$ 
unless an excitation gap opens in the charge degree of freedom,
whereas $N(\vec q)\propto |\vec q|^2$ if a charge gap opens. 
Figure \ref{fig:03CQKSQK}(a) shows $N(\vec q)$ of $\Psi_{\rm AR(opt)}$; 
the behavior for small values of $|\vec q|$ abruptly changes 
between $U/t=8.25$ and 8.3, which again coincide with $U_{\rm c}/t$ 
determined by other quantities. 
For $U>U_{\rm c}$, the behavior becomes $|\vec q|^2$-like, 
suggesting a charge gap opens. 
\par 

Third, we study the spin correlation function, 
\begin{equation} 
S({\vec q})=\frac{1}{N_{\rm s}}\sum_{ij}{e^{i{\vec q}
\cdot({\vec r}_i-{\vec r}_j)} 
\left\langle{S_{i}^zS_{j}^z}\right\rangle}. 
\label{eq:sq}
\end{equation} 
In Fig.~\ref{fig:03CQKSQK}(b), $S({\vec q})$ of $\Psi_{\rm AR(opt)}$ 
is plotted for the same range of $U/t$ as in (a). 
For small $|\vec q|$, the behavior is always 
$S({\vec q})\propto|q|$, suggesting the low-lying spin excitation 
is gapless both in metallic and insulating phases. 
On the other hand, the magnitude at the AF nesting vector 
$\vec q=(\pi,\pi)$ increases as $U/t$ increases, in particular, 
markedly at $U/t=8.3$. 
Thus, the D-H binding wave function shows strong inclination 
toward the antiferromagnetic order, especially in the insulating 
case. 
This aspect is the same as the short-range D-H binding wave 
functions.\cite{YOT} 
\par 

\subsection{System-size dependence of critical point\label{sec:MTSD}}
In this subsection, we consider the system-size dependence 
of the Mott critical point, which we have disregarded to this point. 
\par

\begin{figure}[hob]
 \begin{center}
  \includegraphics[width=75mm,clip]{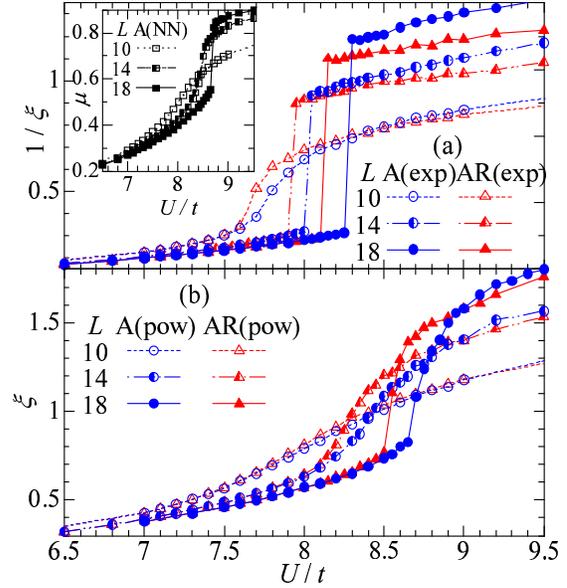}
 \end{center}
 \vskip -3mm
\caption{(Color online) 
The behavior of optimized variational parameters which controls D-H 
binding strength is compared among three system sizes 
near the Mott critical points. 
The results for three types of wave functions are shown: 
(a) $1/\xi$ for A(exp) and AR(exp). 
The inset shows $\mu$ for A(NN). 
(b) $\xi$ for A(pow) and AR(pow).
}
 \label{fig:04DHp}
\end{figure}
%
The optimized variational parameters controlling the D-H binding 
strength is important to definitely determine the critical values. 
In the main panel of Fig~\ref{fig:04DHp}(a), the optimized $1/\xi$ 
in the wave functions of the exponentially-decaying type 
[eq.~(\ref{eq:f(A)}a)] is plotted for three system sizes. 
For small systems like $L=10$, the variation of $1/\xi$ is smooth 
and clear critical behavior is not seen, whereas systems 
with $L\ge 14$ exhibit clear discontinuities. 
Such tendency is often observed when finite systems are used,\cite{HM} 
because the phase transition is well defined for $L=\infty$. 
As another example, we show the result for the short-range D-H 
correlation in the inset of Fig.~\ref{fig:04DHp}(a).\cite{YOT,YMO} 
\par 

It is also a general tendency of the D-H binding wave functions that 
the Mott critical value increases as $L$ increases.\cite{YOT,YMO} 
This stems from the great system-size dependence in $E/t$ 
in the insulating side of the transition point, compared with 
in the metallic side. 
This is reflected in some decrease of the critical value by adding 
repulsive Jastrow factors [$\Psi_{\rm AR(exp)}$], which improve 
$E/t$ especially in the insulating side, as in the inset 
of Fig.~\ref{fig:02EU_Uc}. 
\par

\begin{table}[hob]
\caption{
The Mott critical values, $U_{\rm c}/t$, estimated from eq.~(\ref{eq:Uc}) 
is entered in the first line of each wave function. 
The results of five system sizes are compared. 
The figures in the second line represent the critical values determined 
by the crossing point of $\ell_{\rm DH}$ and $\ell_{\rm DD}$, which 
will be discussed in \S\ref{sec:mecha}. 
The figures with * for A(bind) are estimated from $Z$, which will be 
explained in \S\ref{sec:mecbind}. 
}
\label{t1}
\begin{center}
\begin{tabular}{c|c|c|c|c|c}
\hline
$\Psi$   & $L=10$ & $L=12$ & $L=14$ & $L=16$ & $L=18$ \\
\hline\hline
 A(NN) &  7.85  &  8.25  & 8.475  &  8.575 & 8.675 \\
$\ell_{\rm DH}=\ell_{\rm DD}$ & 8.85 & 8.725 & 8.575 & 8.575 & 8.875 \\ 
\hline
 A(exp)  & 7.75   & 7.875  & 8.025  & 8.125  & 8.275 \\
$\ell_{\rm DH}=\ell_{\rm DD}$ & 7.95 & 7.875 & 8.025 & 8.125 & 8.275 \\ 
\hline
 AR(exp) & 7.65   & 7.775  & 7.925  & 8.025  & 8.125 \\
$\ell_{\rm DH}=\ell_{\rm DD}$ & 7.75 & 7.775 & 7.925 & 8.025 & 8.125 \\ 
\hline
 A(pow)  & 7.55   & 7.95   & 8.375  & 8.575  & 8.675 \\
$\ell_{\rm DH}=\ell_{\rm DD}$ & 8.55 & 8.475 & 8.525 & 8.625 & 8.725 \\ 
\hline
 AR(pow) & 7.65   & 7.95   & 8.175  & 8.375  & 8.525 \\
$\ell_{\rm DH}=\ell_{\rm DD}$ & 8.35 & 8.275 & 8.325 & 8.425 & 8.525 \\
\hline
 A(opt)  & 7.15   & 8.075  & 8.275  & 8.525  & 8.875 \\
$\ell_{\rm DH}=\ell_{\rm DD}$ & 8.75 & 8.775 & 8.825 & 8.925 & 9.025 \\
\hline
 AR(opt) & 7.45   & 7.925  & 8.125  & 8.275  & 8.375 \\
$\ell_{\rm DH}=\ell_{\rm DD}$ & 8.225 & 8.125 & 8.175 & 8.275 & 8.375 \\
\hline\hline
 A(bind) & 2.10*  &  2.28* & 2.365* &  2.39* & 2.42* \\
$\ell^*_{\rm DD}=3$ & 2.00 & 2.175 & 2.275 & 2.375 & 2.425 \\
\hline\hline
\end{tabular}
\end{center}
\label{table:Uc}
\end{table}
%

These tendencies are common to other D-H binding wave functions; 
as another example, in Fig.~\ref{fig:04DHp}(b), we show optimized 
$\xi$ in the power-law decaying correlation factor eq.~(\ref{eq:f(A)}b). 
In $\Psi_{\rm A(pow)}$, the critical behavior is less clear. 
In case the Mott critical point cannot be clearly determined, 
we estimate $U_{\rm c}/t$ from the maximum of the decreasing 
rate of doublon density $d$, 
\begin{equation}
-\frac{\partial d}{\partial (U/t)}
=-\frac{d[(U + \Delta U)/t] - d[U/t]}{\Delta U/t}. 
\label{eq:Uc}
\end{equation}
The Mott critical values thus obtained are summarized 
in Table \ref{table:Uc}. 
\par

\subsection{Effects of different D-H binding factors\label{sec:D-H}}
\begin{figure}[hob]
 \begin{center}
  \includegraphics[width=80mm,clip]{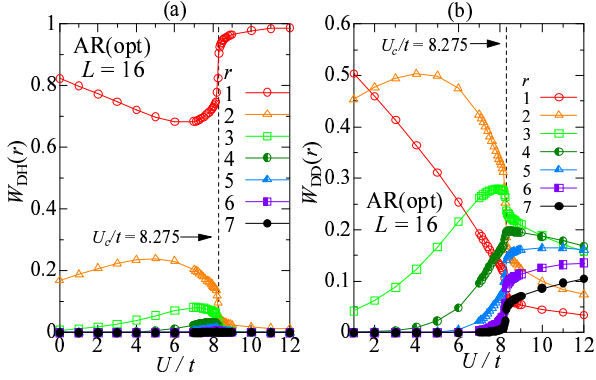}
 \end{center}
 \vskip -3mm
\caption{(Color online)
The appearance rates of nearest (a) D-H and (b) D-D distances of $r$ 
are plotted for $r=1$-7 as a function of $U/t$. 
Both $W_{\rm DH}$ and $W_{\rm DD}$ are calculated with AR(opt). 
The vertical dashed lines indicate the Mott critical point.
}
\label{fig:07PddPdh_multidh}
\end{figure}
%
First of all, we look at the distribution of distance from a doublon 
(holon) to the nearest holon (doublon), which is denoted simply 
by $r$, here. 
We indicate the appearance rate of $r$ by $W_{\rm DH}(r)$, 
which satisfies,
\begin{equation} 
\sum_{r=1}^LW_{\rm DH}(r)=1. 
\end{equation}
In Fig.~\ref{fig:07PddPdh_multidh}(a), $W_{\rm DH}(r)$ with $r\le 7$ 
for the best function $\Psi_{\rm AR(opt)}$ is depicted versus $U/t$. 
In the metallic regime, $W_{\rm DH}(1)$ is predominant, but $W_{\rm DH}(r)$ 
with $r\ge 2$ has an appreciable weight especially at $U/t\sim 6$; there, 
a doublon is detached from holons to some extent. 
Meanwhile, in the insulating regime, the weight is almost concentrated 
on $r=1$, indicating D-H pairs are confined within mutually NN sites. 
In a similar way, we define $W_{\rm DD}(r)$ as the appearance rate 
of doublons (holons) with the D-to-D (H-to-H) distance of $r$. 
As shown in Fig.~\ref{fig:07PddPdh_multidh}(b), $W_{\rm DD}(r)$ of large 
$r$ increases and $W_{\rm DD}(1)$ and $W_{\rm DD}(2)$ rapidly decreases,
as $U/t$ increases. 
In the insulating phase, doublons tend to keep away from each other. 
\par

\begin{figure}[hob]
 \begin{center}
  \includegraphics[width=75mm,clip]{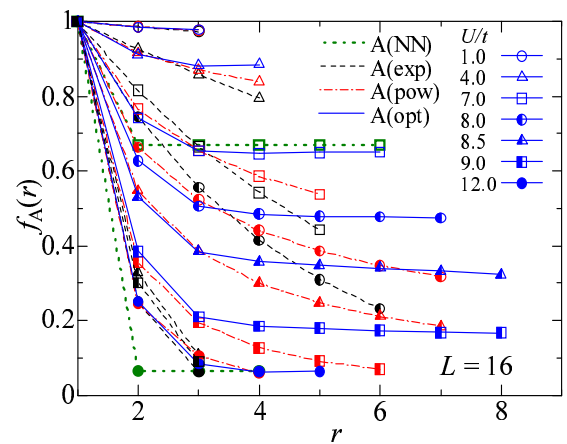}
 \end{center}
 \vskip -5mm
\caption{(Color online) 
Comparison of optimized D-H attractive correlation weight $f_{\rm A}(r)$ 
among four D-H binding wave functions, A(NN), A(exp), A(pow) and A(opt), 
for various values of $U/t$. 
We omit the data which do not meet the condition (22).
}
 \label{fig:05Cdh}
\end{figure} 
%
In the following, we discuss the effects of different D-H binding 
factors $P_{\rm A}^{({\rm x})}$ (x~= a,~b,~c,~NN), without introducing 
repulsive correlations. 
Figure \ref{fig:05Cdh} compares the optimized weights $f_{\rm A}(r)$ 
[eq.~(\ref{eq:f(A)})] among the four D-H projectors. 
The plotted data are restricted to what satisfies the following 
condition: 
\begin{equation} 
\rho(r)>\rho_{\rm min}, 
\label{eq:rho-cond}
\end{equation} 
where $r$ has the same meaning as in the preceding paragraph, 
and $\rho(r)$ [$=W_{\rm DH}(r)\times d$ ] denotes the appearance 
probability of the doublon (holon) of $r$. 
After a search, we put $\rho_{\rm min}=4\times10^{-4}$, which 
corresponds to the probability that doublons or holons of certain $r$ 
appear $10^4$ times in $2.5\times 10^5$ samples for the system of $L=10$.
If the condition (\ref{eq:rho-cond}) is not satisfied for $r=r^*$, 
the optimized $f_{\rm A}(r^*)$ become statistically unreliable 
in the present calculations, and corresponding samples actually make 
only an imperceptible contribution to the averages. 
\par 

We start with the analysis of the optimizing-type wave function 
$\Psi_{\rm A(opt)}$. 
For any values of $U/t$, $f_{\rm A}(r)$ rapidly decreases for $r\lsim 3$, 
but becomes almost constant for $r\gsim 3$. 
Note that this long-range behavior of $f_{\rm A}(r)$ is convenient 
for the conductive nature in the metallic regime ($U<U_{\rm c}$), 
namely, a doublon is released from the bondage of holons, once 
a doublon goes three lattice constants away from holons. 
The effective range of $f_{\rm A}(r)$ satisfying the condition 
(\ref{eq:rho-cond}) becomes the widest ($r\lsim 8$) near $U_{\rm c}/t$, 
but abruptly shrinks in the insulating phase, as expected from 
the data in Fig.~\ref{fig:07PddPdh_multidh}(a). 
In addition, for $U>U_{\rm c}$, the magnitude of $f_{\rm A}(r)$ 
itself is small. 
As a result, a doublon and a holon come to confine each other 
in a narrow D-H pair domain.  
We will return to this topic as to the mechanism of Mott transitions 
in \S\ref{sec:mecha}. 
The propriety of other $\Psi_{\rm A}$ depends on how properly $f_{\rm A}(r)$ of 
$\Psi_{\rm A}$ can imitate that of $\Psi_{\rm A(opt)}$. 
\par

Regarding a short-range D-H factor, the reason why simple $\Psi_{\rm A(NN)}$ 
is unexpectedly good for $U<U_{\rm c}$ (inset of Fig.~\ref{fig:02EU_Uc}) 
is that $f_{\rm A}(r)$ of $\Psi_{\rm A(NN)}$ is constant for $r\ge 2$ 
and resembles that of $\Psi_{\rm A(opt)}$ except for $f_{\rm A}(2)$, 
as shown in Fig.~\ref{fig:05Cdh} for the data of $U/t=7$. 
On the other hand, the reason why $E/t$ of $\Psi_{\rm A(NN)}$ is 
relatively high in the insulating regime, as compared with the other 
$\Psi_{\rm A}$, stems from an underestimate of $f_{\rm A}(2)$, 
as seen for $U/t=12$. 
\par

We turn to the long-range D-H factors. 
In the metallic regime, $f_{\rm A}(r)$'s of $\Psi_{\rm A(pow)}$ and 
especially of $\Psi_{\rm A(exp)}$ are underestimated for small $r$ and 
overestimated for large $r$, to be optimized as a whole. 
Consequently, $\Psi_{\rm A(exp)}$ has appreciably higher energy than 
other $\Psi_{\rm A}$'s; this behavior lowers the Mott critical value 
of $\Psi_{\rm A(exp)}$. 
On the other hand, in the insulating regime, $f_{\rm A}(r)$'s of 
both $\Psi_{\rm A(exp)}$ and $\Psi_{\rm A(pow)}$ almost coincide with 
that of $\Psi_{\rm A(opt)}$, because the long-range part of $f_{\rm A}(r)$ 
($r\gsim 3$) substantially vanishes. 
As a result, $E(\Psi_{\rm A(exp)})$ and $E(\Psi_{\rm A(pow)})$ become 
as good as $E(\Psi_{\rm A(opt)})$. 
\par 

\subsection{Effects of repulsive intersite correlations\label{sec:repulsive}}
First of all, we compare contributions to the improvement of energy 
between D-H attractive and D-D (and H-H) repulsive correlation factors.  
In Table \ref{table:optE}, the optimized total energies of GWF and 
the optimizing-type wave functions eq.~(\ref{eq:Psiopt}) are summarized. 
The results of other types of wave functions eqs.~(\ref{eq:Psiexp}) 
and (\ref{eq:Psipow}) are basically identical as far as the effect 
of repulsive factors is concerned. 
The wave function $\Psi_{\rm R(opt)}$ [eq.~(\ref{eq:Psiopt})], 
in which only the 
repulsive factor $P_{\rm R}$ [eq.~(\ref{eq:PR(jas)}) with 
eq.~(\ref{eq:f(R)}c)] is applied to GWF, improves $E/t$ only very slightly 
on GWF for any value of $U/t$. 
Furthermore, $\Psi_{\rm R(opt)}$ never induces a Mott transition, 
like GWF. 
Thus, the repulsive correlation $P_{\rm R}$, by itself, make no 
substantial improvement on GWF. 
In contrast, as already discussed for Table \ref{table:Uc} and 
Fig.~\ref{fig:02EU_Uc}, $\Psi_{\rm A(opt)}$ induces a Mott transition, 
and make a great improvement in $E/t$ on GWF for $U\gsim U_{\rm c}$, 
conclusively showing that the essence of Mott transition is included 
in the D-H binding correlation. 
$\Psi_{\rm AR(opt)}$ ($=P_{\rm R}\Psi_{\rm A(opt)}$) further reduces 
$E/t$ for $U>U_{\rm c}$. 
Although the decrement in energy thereby is as small as 2\% of 
$E(\Psi_{\rm A(opt)})$ [$U/t=9$ and 12], the addition of $P_{\rm R}$ 
shifts the Mott critical point to a somewhat lower value 
(Table \ref{table:Uc}), and make the critical behavior more 
first-order-like (inset of Fig.~\ref{fig:03ZU_Uc}). 
\par

\begin{table}[hob]
\caption{
Comparison of variational energies $E/t$ among GWF and the optimizing 
type of wave functions, eq.~(\ref{eq:Psiopt}), for $L=16$. 
The digits in the brackets denotes the error in the last digits. 
}
\label{t1}
\begin{center}
\begin{tabular}{c|c|c|c|c}
\hline
$U/t$ &   GWF    & R(opt) &   A(opt)   &  AR(opt)   \\
\hline\hline
1.0  & -1.3865(2) & -1.3866(3) & -1.3867(4) & -1.3867(3) \\
\hline
7.0  & -0.374(1)  & -0.375(2)  & -0.418(2)  & -0.418(2) \\
\hline
7.5  & -0.323(2)  & -0.323(2)  & -0.379(2)  & -0.379(2) \\
\hline
8.0  & -0.276(2)  & -0.277(2)  & -0.348(2)  & -0.348(3) \\
\hline
8.5  & -0.234(2)  & -0.234(2)  & -0.325(3)  & -0.330(2) \\
\hline
9.0  & -0.196(2)  & -0.197(2)  & -0.310(2)  & -0.317(2) \\
\hline
12.0 & -0.061(2)  & -0.063(2)  & -0.255(3)  & -0.261(2) \\
\hline\hline
\end{tabular}
\end{center}
\label{table:optE}
\end{table}
%
\begin{figure}[hob]
 \begin{center}
  \includegraphics[width=70mm,clip]{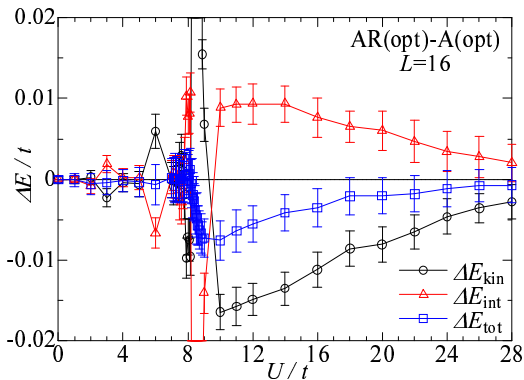}
 \end{center}
 \vskip -3mm
\caption{(Color online) 
The differences of total, kinetic and interaction energies between  
AR(opt) and A(opt) are plotted as a function of $U/t$ for $L=16$. 
If the value is negative, AR(opt) has a lower energy than A(R). 
The large deviations near $U_{\rm c}/t$ should be neglected, which 
stem from the discordance of $U_{\rm c}/t$. 
}
 \label{fig:05optddhh_dh}
\end{figure}
%
Next, we consider the differences of kinetic and of interaction 
energies between $\Psi_{\rm AR(opt)}$ and $\Psi_{\rm A(opt)}$: 
\begin{equation}
\Delta E_\Gamma=E_\Gamma[\mbox{AR(opt)}]-E_\Gamma[\mbox{A(opt)}]
\label{eq:delE}
\end{equation}
where the suffix $\Gamma$ denotes ``kin", ``int" or ``tot". 
In Fig~.\ref{fig:05optddhh_dh}, the three kinds of $\Delta E$ are 
plotted versus $U/t$. 
In the metallic regime, not only $\Delta E_{\rm tot}$ remains zero, 
but also both $\Delta E_{\rm kin}$ and $\Delta E_{\rm int}$ are zero 
except for irregular accidental deviations. 
Thus, $P_{\rm R}$ does not modify $E(\Psi_{\rm A(opt)})$ for $U<U_{\rm c}$. 
Nevertheless, for $U/t>10$ in the insulating regime, $\Delta E_{\rm kin}$ 
exhibits appreciable negative values, according to $\Delta E_{\rm tot}$. 
Inversely, $\Delta E_{\rm pot}$ has regular positive values; 
$P_{\rm R}$ with the aid of $P_{\rm A}$ reduces $E_{\rm kin}$ 
at the cost of $E_{\rm int}$ in the insulating phase. 
Thus, $P_{\rm R}$ compensates for the excess of D-H binding effects. 
\par

\begin{figure}[hob]
 \begin{center}
  \includegraphics[width=70mm,clip]{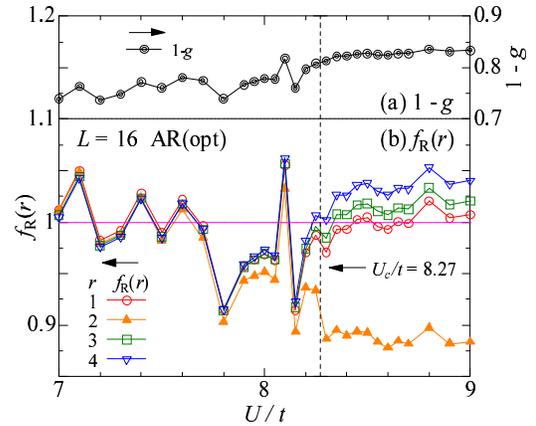}
 \end{center}
 \vskip -3mm
\caption{(Color online) 
(a) Optimized onsite repulsive (Gutzwiller) parameter $g$ near the Mott 
critical point. 
(b) Optimized long-range repulsive correlation weight $f_{\rm R}(r)$, 
eq.~(\ref{eq:f(R)}c). 
The results in (a) and (b) are obtained in an identical calculation 
for AR(exp). 
}
\label{fig:05Cdd}
\end{figure}
%
Finally, we discuss the repulsive correlation weight for 
$\Psi_{\rm AR(opt)}$, in which $f_{\rm R}(r)$ is optimized 
for each $r$. 
In Fig.~\ref{fig:05Cdd}(b), $f_{\rm R}(r)$ for $r\le 4$ is shown 
near $U_{\rm c}/t$.  
In the metallic regime, $f_{\rm R}(r)$ is almost unity, 
although there is some fluctuations, indicating the repulsive intersite 
correlation is virtually ineffective. 
Correspondingly, $\Psi_{\rm AR(opt)}$ seldom improves $E/t$ on that 
of $\Psi_{\rm A(opt)}$ for $U<U_{\rm c}$ (Table \ref{table:optE}). 
Meanwhile, in the insulating regime, only $f_{\rm R}(2)$ is slightly 
lowered from unity, causing the sudden drop of $W_{\rm DD}(2)$ 
for $U>U_{\rm c}$ [Fig.~\ref{fig:07PddPdh_multidh}(b)].\cite{noteWDD} 
The behavior of $f_{\rm R}(r)>1$ for $r\ge 3$ indicates that D-D and H-H 
correlations are {\it attractive} rather than repulsive 
in a long-range part.\cite{notefR} 
This behavior promotes the repulsion between doublons in proximity, 
as seen in Fig.~\ref{fig:07PddPdh_multidh}(b), where $W_{\rm DD}(r)$ 
for $r>3$ is larger than $W_{\rm DD}(1)$. 
These effects of $f_{\rm R}(r)$ causes a small but steady improvement 
in energy on $\Psi_{\rm A(opt)}$ (Table \ref{table:optE}). 
Incidentally, the scattered data, especially for $U<U_{\rm c}$, stem 
from the complementarity between the onsite and intersite repulsive 
correlations. 
The behavior of $g$ in Fig.~\ref{fig:05Cdd}(a) is quite opposite 
to that of $f_{\rm R}(r)$ in Fig.~\ref{fig:05Cdd}(b). 
Owing to the balance between the two, the resultant physical quantities 
become smooth, for instance, $d$ in Fig.~\ref{fig:07Cmultidh}. 
In this point, the parameter space has redundancy. 
\par

\section{Improved Picture of Mott Transitions\label{sec:mecha}}
In \S\ref{sec:mecbind}, we discuss the Mott transition arising 
in the completely D-H bound state, and provide a reformed picture 
of Mott transitions to comprehend it. 
In \S\ref{sec:ordinary}, we show that this picture is applicable 
to various D-H binding wave functions. 

\subsection{Mott transition in completely D-H bound state\label{sec:mecbind}}
\begin{figure}[hob]
 \begin{center}
  \includegraphics[width=80mm,clip]{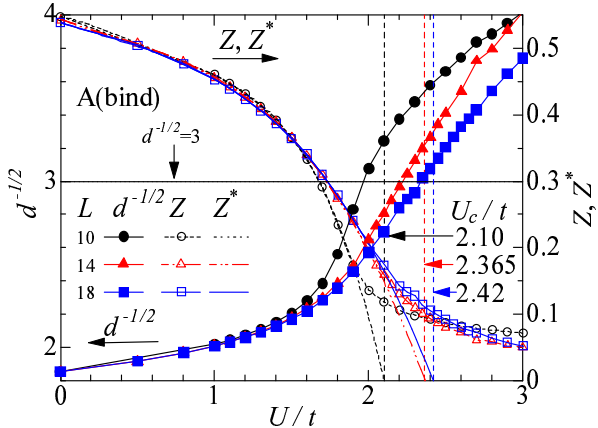}
 \end{center}
 \vskip -3mm
\caption{(Color online) 
The Mott critical value $U_{\rm c}/t$ for completely D-H bound state 
A(bind) is estimated in two ways.
The vertical dashed lines indicate the values estimated from the 
extrapolation ($Z^*$) to zero of the quasiparticle renormalization 
factor $Z$. 
The other estimate is given by the crossing point of $\ell_{\rm DH}^*$ 
($=3$) and $\ell_{\rm DD}^*$ ($=1/\sqrt{d}$). 
The results are listed in Table.~\ref{table:Uc}. 
}
\label{fig:06Cmu1}
\end{figure}

In \S\ref{sec:energy}, we supposed that the completely D-H bound 
wave function $\Psi_{\rm A(bind)}$ [eq.~(\ref{eq:FA(bind)})] 
represents a typical insulating state, but this supposition is 
not true for small values of $U/t$. 
The total energy of $\Psi_{\rm A(bind)}$ in Fig.~\ref{fig:02EU_Uc} 
exhibits an abrupt change of curvature at $U/t\sim 2.5$. 
Corroboratively, as shown in Fig.~\ref{fig:03ZU_Uc}, the quasiparticle 
renormalization factor $Z$ has finite values for $U/t\lsim 2.5$. 
Thus, it is certain that $\Psi_{\rm A(bind)}$ exhibits a Mott transition 
and is metallic for small $U/t$. 
Then, we estimate the Mott critical values of $\Psi_{\rm A(bind)}$ 
from the extrapolation of $Z$ for $U/t\lsim 2$ using the third-order 
least squares method, as shown in Fig.~\ref{fig:06Cmu1}, and summarize 
them in Table \ref{table:Uc}. 
In $\Psi_{\rm A(bind)}$, doublons (holons) must be accompanied by 
at least one holon (doublon) in its NN sites. 
The metallic state satisfying this condition cannot be represented 
simply by the unbinding of D-H pairs proposed in ref.~\citen{YOT}.
In what follows, we consider a microscopic picture of Mott transitions 
which comprehends the case of $\Psi_{\rm A(bind)}$. 
\par

%
For simplicity, we take up a one-dimensional case,\cite{note1D} 
and first assume that $U/t$ is small and the doublon (and holon) density 
is sufficiently high. 
Actually, for $U/t\lsim 1.5$, $d$ of $\Psi_{\rm A(bind)}$ becomes 
higher than those of the other D-H binding states [see Fig.~\ref{fig:02EkinD}(b)]. 
Then, a doublon are often accompanied by multiple holons in the NN sites, 
so that it can propagate independently of holons as a plus charge carrier, 
satisfying the complete binding condition: 
\begin{eqnarray}
\nonumber &&
[\cdot\cdot {\rm DH\dot D}_\leftrightarrow{\rm \dot HDH}\cdot\cdot]
\rightarrow
[\cdot\cdot {\rm DH}_\leftrightarrow{\rm \dot H\dot D}_\leftrightarrow{\rm DH} \cdot\cdot]
\rightarrow \qquad
\\
&&
[\cdot\cdot {\rm D}_\leftrightarrow{\rm \dot HHD\dot D}_\leftrightarrow{\rm H} \cdot\cdot]
\rightarrow
[\cdot\cdot_\leftrightarrow{\rm \dot HDHDH\dot D}_\leftrightarrow\cdot\cdot]
\rightarrow \qquad. 
\label{eq:DHDH}
\end{eqnarray}
Here, pay attention to the dotted D and H, which propagate by exchanging 
the positions according to the double-headed arrows. 
Thus, the state becomes metallic. 
On the other hand, when $U/t$ becomes large and the doublon density 
becomes sufficiently low, most D-H pairs become mutually detached, as 
$
[\cdot\cdot {\rm \uparrow\downarrow HD\downarrow\uparrow HD\uparrow\downarrow}\cdot\cdot]. 
$
Then, a doublon becomes unable to propagate independently of holons, 
and vice versa, satisfying the completely bound condition. 
Consequently, charge fluctuation is confined locally, resulting in 
an insulator. 
Summing up, the conduction in a metallic state requires that D-H pairs 
should contact with one another in sequence; the Mott critical value 
can be specified by $U/t$ at which D-H pairs are mutually detached. 
\par

From the above argument, we find it convenient for general discussions 
of Mott transitions to introduce two characteristic length scales, 
the D-H binding length $\ell_{\rm DH}$ and the D-D exclusion length 
$\ell_{\rm DD}$, which are generally a function of $U/t$. 
We postulate that the attractive correlation factor $P_{\rm A}$ 
produces D-H pairs of a binding length $\ell_{\rm DH}$ according 
to $U/t$; $\ell_{\rm DH}$ approximately corresponds to the size 
of a D-H-pair domain, in which at least one doublon and one holon 
must exist. 
$\ell_{\rm DD}$ broadly represents the distance between two D-H pairs. 
Using $\ell_{\rm DH}$ and $\ell_{\rm DD}$, we give a microscopic 
picture of Mott transitions, which is schematically shown 
in Figs.~\ref{fig:mec2}(c) and (d) for general D-H binding wave functions. 
In the insulating phase, the relation $\ell_{\rm DH}<\ell_{\rm DD}$ 
holds, indicating that the domains of D-H pairs do not usually overlap, 
at least, not in sequence.  
Consequently, most D-H pairs are isolated and a doublon and a holon 
are confined within $\ell_{\rm DH}$, resulting in only local charge 
fluctuation. 
To this point, the picture is basically identical with the previous 
one.\cite{YOT} 
In the conductive phase ($U<U_{\rm c}$), $\ell_{\rm DH}$ becomes 
larger than $\ell_{\rm DD}$, indicating the domains of D-H pairs 
overlap with one another. 
Then, a doublon in a D-H pair can exchange a partner holon with a holon 
in an adjacent D-H pair, when the two holons are in the overlapped 
area. 
As a result, a doublon and a holon can move independently as charge 
carriers by successively exchanging the partner. 
As $U/t$ is varied, a Mott transition takes place when $\ell_{\rm DH}$ 
becomes equivalent to $\ell_{\rm DD}$, 
which is roughly $1/\sqrt{d}$ and generally a monotonically increasing 
function of $U/t$. 
In this framework, it is of primarily important to determine 
$\ell_{\rm DH}$ and $\ell_{\rm DD}$ appropriately. 

\par

\begin{figure}[hob]
 \vskip -3mm 
 \begin{center}
  \includegraphics[width=50mm,clip]{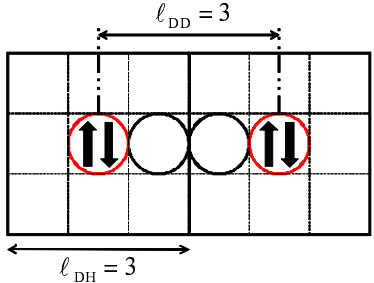}
 \end{center}
 \vskip -2mm 
\caption{(Color online) 
Two domains of D-H pairs for the completely bound state A(bind) 
(domain size $\ell_{\rm DH}$ is 3) are schematically shown
at the Mott critical case: $\ell_{\rm DD}=\ell_{\rm DH}$. 
Doublons and holons are represented by circles with and without 
arrows, respectively. 
The empty squares indicate singly occupied sites. 
}
\label{fig:06mu1mec}
\end{figure}
%
Now, we apply the above framework to the special case of 
$\Psi_{\rm A(bind)}$. 
We postulate that a domain of a D-H pair for $\Psi_{\rm A(bind)}$ 
consists of $3\times 3$ lattice sites, as shown in Fig.~\ref{fig:06mu1mec}. 
Note that the size of this domain for $\Psi_{\rm A(bind)}$ is constant, 
irrespective of $U/t$, because a doublon and a holon (or holons) are 
tightly bound in the nearest neighbor site(s). 
For $\ell_{\rm DH}<3$, some domains of D-H pairs mutually overlap, 
and sequence like DHDHDH in (\ref{eq:DHDH}) possibly appears, 
whereas for $\ell_{\rm DH}>3$ the domains of D-H pairs separate 
from  each other on average. 
Thus, it seems appropriate to put $\ell_{\rm DH}^*=3$. 
Here, we add a star to distinguish it from the form for ordinary 
(not completely bound) $\Psi_{\rm A}$'s discussed later. 
As for the D-D exclusion length, we simply put $\ell_{\rm DD}^*=1/\sqrt{d}$, 
as mentioned above. 
In Fig.~\ref{fig:06Cmu1}, we plot $\ell_{\rm DD}^*$ versus $U/t$. 
The values of $U/t$ at which $\ell_{\rm DH}^*=\ell_{\rm DD}^*$ are 
summarized in Table.~\ref{table:Uc}, and approximately consistent 
with the Mott critical values estimated from $Z$. 
Thus, the above picture with a broad estimate of $\ell_{\rm DH}$ and 
$\ell_{\rm DD}$ yields a justifiable result to this special case. 
\par


\subsection{Application to ordinary D-H wave functions\label{sec:ordinary}}
\begin{figure}[hob]
 \begin{center}
  \includegraphics[width=80mm,clip]{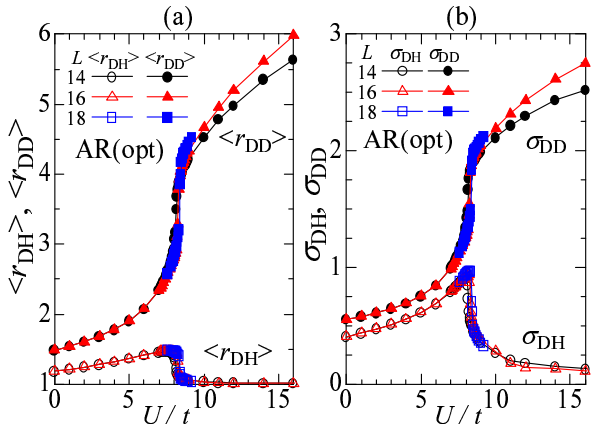}
 \end{center}
 \vskip -3mm
\caption{(Color online)
(a) The nearest D-to-H and H-to-D distances and the nearest 
D-to-D and H-to-H distances calculated with AR(opt) are plotted 
for three system sizes as a function of the interaction strength.  
(b) Standard deviations of $\langle r_{\rm DH}\rangle$ and 
of $\langle r_{\rm DD}\rangle$ shown in (a). 
}
\label{fig:07Rmultidh}
\end{figure}
\begin{figure}[hob]
 \begin{center}
  \includegraphics[width=75mm,clip]{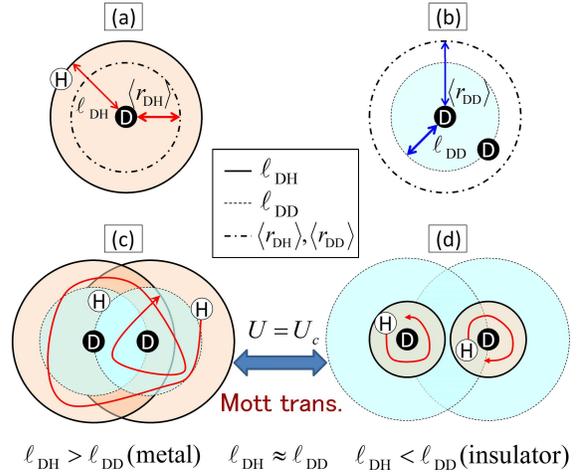}
 \end{center}
 \vskip -2mm
\caption{(Color online) 
Schematic figures of microscopic mechanism of the Mott transition.
(a) The solid circle of radius $\ell_{\rm DH}$ denotes the domain 
in which a holon can itinerate, when the partner doublon is located 
at the center. 
(b) The solid circle of radius $\ell_{\rm DD}$ denotes the forbidden 
area where a doublon cannot enter, when another doublon is situated 
at the center. 
In (c) and (d), the behavior of doublons (D) and holons (H) is 
illustrated for metallic and insulating states, respectively, 
in ordinary D-H binding wave functions. 
The explanation is given in \S\ref{sec:mecbind}. 
The two phases are distinguished by comparing the magnitude of 
$\ell_{\rm DH}$ and $\ell_{\rm DD}$. 
}
\label{fig:mec2}
\end{figure}

%
Here, we study whether the above picture of Mott transitions is 
applicable to ordinary D-H binding (not rigidly bound) wave functions 
$\Psi_{\rm A}$ and to those with repulsive factors $\Psi_{\rm AR}$. 
In contrast to $\Psi_{\rm A(bind)}$, the distance from a doublon to 
its nearest holon(s), $\langle r_{\rm DH}\rangle$, varies in the 
ordinary $\Psi_{\rm A}$ and $\Psi_{\rm AR}$, so that not only 
$\ell_{\rm DD}$ but also $\ell_{\rm DH}$ should depend on $U/t$. 
In Fig.~\ref{fig:07Rmultidh}(a), we show the behavior of the averages 
of the nearest D-to-H and H-to-D distances $\langle r_{\rm DH}\rangle$ 
and of the nearest D-to-D and H-to-H distances $\langle r_{\rm DD}\rangle$ 
calculated with $\Psi_{\rm AR(opt)}$ are plotted as a function of $U/t$. 
Figure \ref{fig:07Rmultidh}(b) shows the standard deviations of 
$\langle r_\Lambda\rangle$ ($\Lambda=$ DH or DD), 
\begin{equation}
\sigma_{\Lambda} = \sqrt{\frac{1}{M}
\sum_{i=1}^{M}(r_i-\langle r_{\Lambda}\rangle)^2}, 
\label{eq:SD} 
\end{equation}
where the index $i$ runs over all the doublons and holons in all the 
measured samples, and $M$ indicates their total number.  
$\sigma_{\Lambda}$ behaves similarly to $\langle r_\Lambda\rangle$. 
Allowing for the meaning of $\ell_{\rm DH}$ and $\ell_{\rm DH}$, 
we put, generally, 
\begin{eqnarray}
\ell_{\rm DH} &=& \langle r_{\rm DH} \rangle + \sigma_{\rm DH}, 
\label{eq:lDH} \\
\ell_{\rm DD} &=& \langle r_{\rm DD} \rangle - \sigma_{\rm DD}. 
\label{eq:lDD} 
\end{eqnarray}
Namely, $\ell_{\rm DH}$ broadly represents the maximum radius 
of a D-H pair domain, over which a doublon does not separate from holons, 
as in Fig.~\ref{fig:mec2}(a), and $\ell_{\rm DD}$ the minimum D-D 
length, under which doublons do not approach each other, as 
in Fig.~\ref{fig:mec2}(b). 
\par

\begin{figure}[hob]
 \begin{center}
  \includegraphics[width=80mm,clip]{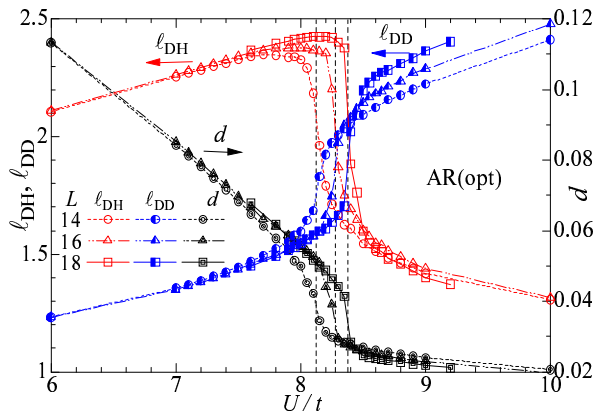}
 \end{center}
 \vskip -3mm
\caption{(Color online) 
The D-H binding length $\ell_{\rm DH}$ and the D-D exclusion distance
$\ell_{\rm DD}$ for AR(opt) obtained from the data 
in Fig.~\ref{fig:07Rmultidh} are plotted as a function of $U/t$. 
For comparison, the doublon density and the critical values obtained 
thereby (vertical dashed lines) are added for the same systems. 
}
\label{fig:07Cmultidh}
\end{figure}

%
In the metallic regime, $\langle r_{\rm DH}\rangle$ and $\sigma_{\rm DH}$ 
gradually increase as $U/t$ increases as shown in Fig.~\ref{fig:07Rmultidh},
chiefly because the densities of doublon and holon are reduced 
by $U/t$ [see Fig.~\ref{fig:02EkinD}(b)]. 
This effect exceeds the D-H binding effect of $P_{\rm A}$. 
Consequently, a doublon somewhat separate from holons. 
At the Mott critical point, however, $\langle r_{\rm DH}\rangle$ 
and $\sigma_{\rm DH}$ suddenly drop, and asymptotically approach 1 and 0, 
respectively, in the insulating regime, owing to the predominant 
D-H binding effect.
In contrast, $\langle r_{\rm DD}\rangle$ and $\sigma_{\rm DD}$ 
monotonically increase as $U/t$ increases, owing to the steady 
decrease of $d$. 
 In Fig.~\ref{fig:07Cmultidh}, $\ell_{\rm DH}$ and $\ell_{\rm DD}$ obtained through 
eqs.~(\ref{eq:lDH}) and (\ref{eq:lDD}) are plotted. 
The Mott critical points $U_{\rm c}/t$ are estimated from the steepest 
descent of an order parameter of the Mott transition $d$, and are 
listed in Table \ref{table:Uc}. 
It is found that $\ell_{\rm DH}$ intersects $\ell_{\rm DH}$ almost 
at $U_{\rm c}/t$. 
Namely, the Mott critical point is estimated also from the condition,
\begin{equation}
\ell_{\rm DH}=\ell_{\rm DD}. 
\label{eq:Mottcondition}
\end{equation}
The values of $U_{\rm c}/t$ thus obtained are listed 
in Table \ref{table:Uc}. 
Thus, the picture of the Mott transition introduced in \S\ref{sec:mecbind} 
[Figs.~\ref{fig:mec2}(c) and \ref{fig:mec2}(d)] is applicable 
to $\Psi_{\rm AR(opt)}$. 
\par

\begin{figure*}[!t]
 \begin{center}
  \includegraphics[width=135mm,clip]{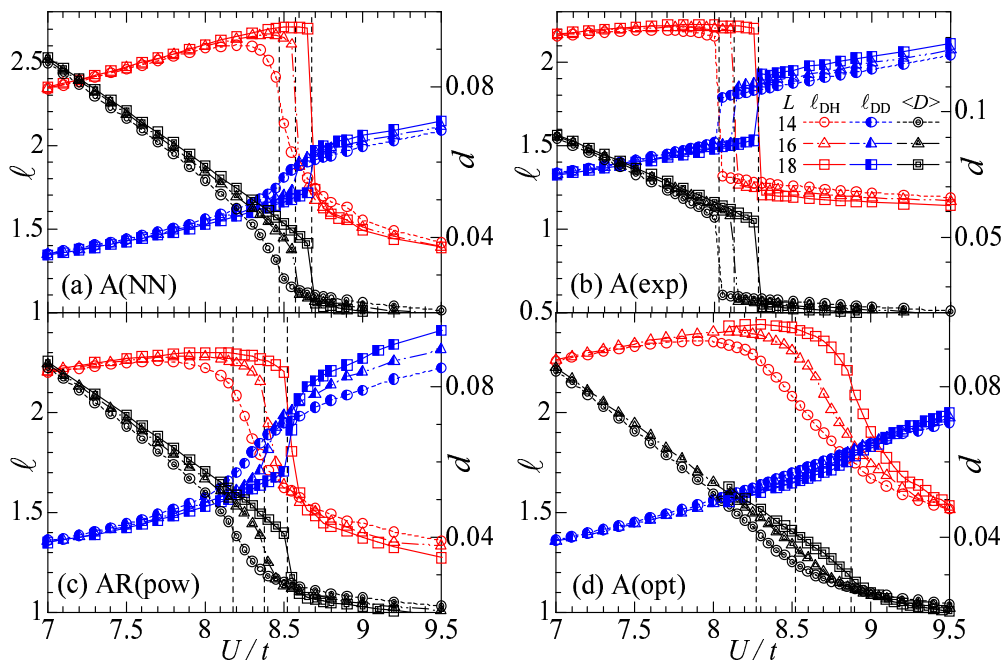}
 \end{center}
 \vskip -3mm
\caption{(Color online) 
The same quantities as in Fig.~\ref{fig:07Cmultidh} ($\ell_{\rm DH}$, 
$\ell_{\rm DD}$, $d$) are shown in the same way for four different 
types of wave functions: (a) A(NN), (b) A(exp), (c) AR(pow) and (d) A(opt). 
The symbols indicated in (b) are common to all panels. 
The Mott critical values $U_c/t$ obtained by $d$ and the crossing point 
of $\ell_{\rm DH}$ and $\ell_{\rm DD}$ are summarized 
in Table.~\ref{table:Uc}. 
}
 \label{fig:08Cother} 
\end{figure*}

%
Now, we confirm whether the above scheme with eqs.~(\ref{eq:lDH}) 
and (\ref{eq:lDD}) are effective also for other wave functions. 
We have made similar analyses for various types of $\Psi_{\rm A}$ 
and $\Psi_{\rm AR}$ in Table \ref{table:index}. 
As a results, the Mott critical values determined under the condition 
eq.~(\ref{eq:Mottcondition}) coincides with $U/t$ estimated from $d$ 
for every wave function with sufficient accuracy at least for large $L$. 
As an example, in Fig.~\ref{fig:08Cother}, we plot the same quantities 
as in Fig.~\ref{fig:07Cmultidh} for four representatives. 
\par

%
\begin{figure}[hob]
 \begin{center}
  \includegraphics[width=65mm,clip]{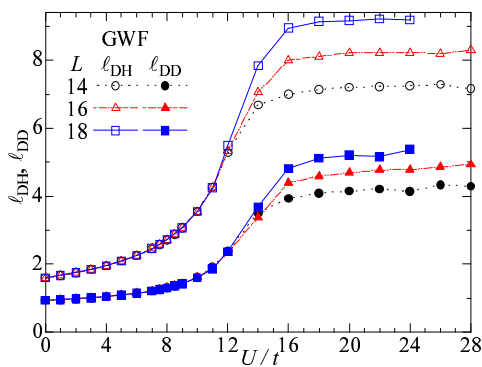}
 \end{center}
 \vskip -3mm 
\caption{(Color online) 
D-H binding length $\ell_{\rm DH}$ and D-D exclusion distance 
$\ell_{\rm DD}$ calculated with Gutzwiller wave function as function 
of interaction strength. 
}
 \label{fig:09CGWF}
\end{figure}
%
Finally, we touch on the wave functions with only repulsive correlation 
factors, namely GWF and the three $\Psi_{\rm R}$'s in Table \ref{table:index}. 
As mentioned in \S\ref{sec:repulsive}, these wave functions do not 
undergo Mott transitions, and are always metallic for $U/t<\infty$. 
In Fig.~\ref{fig:09CGWF}, we plot $\ell_{\rm DH}$ and $\ell_{\rm DD}$ 
of GWF versus $U/t$, as a typical example. 
$\ell_{\rm DD}$ is a monotonically increasing function of $U/t$ 
in the similar way as $\Psi_{\rm A}$ and $\Psi_{\rm AR}$, 
whereas $\ell_{\rm DH}$ is also a monotonically increasing function 
and is always about twice larger than $\ell_{\rm DD}$. 
Namely, the relation $\ell_{\rm DH}>\ell_{\rm DD}$ always holds and 
$\ell_{\rm DH}$ never crosses $\ell_{\rm DD}$. 
The situation is the same for the three $\Psi_{\rm R}$'s. 
Thus, the new picture on the Mott transition properly works for the 
repulsive Jastrow-type wave functions. 
From this argument, it is certified again that the D-H binding effect 
is the essence of the Mott transition. 
\par

\section{Summary\label{sec:summa}}
In this paper, we have studied the nonmagnetic Mott transition 
in the Hubbard model on the square lattice, using a variational Monte 
Carlo method. 
In the trial wave functions, we introduce long-range doublon-holon 
attractive and doublon-doublon (holon-holon) repulsive correlation 
factors, in addition to the onsite repulsive factor. 
We recapitulate the main results below. 
\par

(1) 
We confirmed that the D-H binding correlation is crucial to describe 
first-order nonmagnetic Mott transitions. 
The D-H attractive projector, if properly chosen, considerably reduces 
the variational energy for $U\gsim W$. 
\par 

(2) 
We clarified the optimized weight of the D-H binding factor $f_{\rm A}(r)$, 
which rapidly decreases for $r\lsim 3$, but becomes almost constant 
for $r\gsim 3$ in the metallic regime. 
Thereby, it becomes clear why the simple conventional short-range D-H factors 
capture the essence of the Mott transition. 
In the insulating regime, the D-H binding factor becomes very short-ranged; 
consequently, charge fluctuation, i.e. D-H pairs, is confined within 
$r\lsim 2$. 
\par

(3) 
The D-D repulsive factors improve the energy only slightly, especially 
in the metallic regime, and do not induce a Mott transition by itself. 
However, the improvement in the insulating regime contributes to 
some downward shift of the Mott critical point $U_{\rm c}/t$. 
\par

(4) 
Motivated by the Mott transition in the completely D-H bound state, 
we have renewed the picture of Mott transitions. 
Two characteristic length scales $\ell_{\rm DH}$ and $\ell_{\rm DD}$ 
are introduced; $\ell_{\rm DH}$ broadly represents the size of a D-H pair, 
and $\ell_{\rm DD}$ the minimum distance between two doublons. 
The two lengths generally depend largely on $U/t$, and should be 
appropriately estimated. 
The Mott critical point determined by the condition 
$\ell_{\rm DH}=\ell_{\rm DD}$ is consistent with the values estimated 
from other quantities. 
This picture is applicable to a wide range of Mott transitions including 
the Bose Hubbard model.\cite{YMO}
\par

We leave some intriguing subjects for future studies: 
(i) 
Improvement of the critical point $U_{\rm c}/t$. 
(ii) 
Introduction of explicit AF and superconducting correlation 
into the one-body part.\cite{Tahara} 
(iii) How the new picture works for doped cases, namely, doped 
Mott insulators like the cuprate superconductors.
\par

\begin{acknowledgments}
We would like to thank Masao Ogata for useful discussions. 
This work is partly supported by Grant-in-Aids from 
the Ministry of Education, Culture, Sports, Science and Technology.
\end{acknowledgments}




\begin{thebibliography}{99}


\bibitem{Mott} N.~F.~Mott: {\it Metal-insulator transitions}, 
(Taylor \& Francis, London, 1990).

\bibitem{Greiner} M.~Greiner, O.~Mandel, T.~Esslinger, 
T.~W.~H\"ansch and I.~Bloch: 
\journal{Nature}{415}{39}{2002}. 

\bibitem{1D} T.~St\"oferle, H.~Moritz, C.~Schori, M.~K\"ohl and 
T.~Esslinger: \journal{\PRL}{92}{130403}{2004}.

\bibitem{Kohl} M.~K\"ohl, H.~Moritz, T.~St\"oferle, C.~Schori and 
T.~Esslinger: 
\journal{\JLTP}{138}{635}{2005}. 

\bibitem{Spielman} I.~B.~Spielman, W.~D.~Phillips and J.~V.~Porto: 
\journal{\PRL}{98}{080404}{2007}.

\bibitem{Fisher} For instance, M.~P.~A.~Fisher, P.~B.~Weichman, 
G.~Grinstein and D.~S.~Fisher: 
\journal{\PRB}{40}{546}{1989}. 

\bibitem{Jaksch} D.~Jaksch, C.~Bruder, J.~I.~Cirac, C.~W.~Gardiner 
and P.~Zoller: 
\journal{\PRB}{81}{3108}{1998}.

\bibitem{Bloch-Rev} I.~Bloch, J.~Dalibard and W.~Zwerger: 
\journal{\RMP}{80}{885}{2008}.

\bibitem{critical} At unit filling, the critical value is estimated 
at $U_{\rm c}/t\sim 3.6$ in one dimension (ref.~\citen{Uc-1d}), 
at 16.4-16.7 for the square lattice 
(ref.\citen{QMC-2D1,QMC-2D2,QMC-2D3,Monien}), and at 29.3 
(ref.~\citen{Uc-3d1}) or 31.3 (ref.~\citen{Uc-3d2}) 
for the simple cubic lattice. 

\bibitem{Uc-1d} T.~D.~K\"uhner and H.~Monien: 
\journal{\PRB}{58}{R14741}{1998}.

\bibitem{QMC-2D1} 
W.~Krauth and N.~Trivedi: 
\journal{\EPL}{14}{627}{1991}. 

\bibitem{QMC-2D2}
S.~Wessel, F.~Alet, M.~Troyer and G.~G.~Batrouni: 
\journal{\PRA}{70}{053615}{2004}. 
 
\bibitem{QMC-2D3} 
B.~Capogrosso-Sansone, S.~G.~S\"oyler, N.~Prokof'ev and B.~Svistunov: 
\journal{\PRA}{77}{015602}{2008}. 

\bibitem{Monien} N.~Elstner and H.~Monien: 
\journal{\PRB}{59}{12184}{1999}. 

\bibitem{Uc-3d1} B.~Capogrosso-Sansone, N.~V.~Prokof'ev and 
B.~V.~Svistunov:
\journal{\PRB}{75}{134302}{2007}. 

\bibitem{Uc-3d2} Y.~Kato, Q.~Zhou, N.~Kawashima and N.~Trivedi: 
\journal{\NP}{4}{617}{2008}.

\bibitem{Slater} J.~C.~Slater: \journal{\PR}{82}{538}{1951}.

\bibitem{ET} K.~Kanoda: \journal{Physica C}{282-287}{299}{1997}; 
K.~Miyagawa, K.~Kanoda and A.~Kawamoto: 
\journal{Chem.~Rev.}{104}{5635}{2004}; 
R.~H.~McKenzie: \journal{Science}{278}{820}{1997}. 


\bibitem{PALee}
P.~A.~Lee, N.~Nagaosa and X.~-G.~Wen: \journal{\RMP}{78}{17}{2006}. 

\bibitem{OF} 
M.~Ogata and H.~Fukuyama: \journal{\RPP}{71}{036501}{2008}.

\bibitem{YOTKT} H.~Yokoyama, M.~Ogata, Y.~Tanaka, K.~Kobayashi and 
H.~Tsuchiura: in preparation. 

\bibitem{Hirsch} J.~E.~Hirsch and D.~J.~Scalapino: 
\journal{\PRB}{27}{7169}{1983}.

\bibitem{YS2} H.~Yokoyama and H.~Shiba:
\journal{\JPSJ}{56}{3582}{1987}
\bibitem{Gutz} M.~Gutzwiller: \journal{\PRL}{10}{159}{1963}.

\bibitem{BR} W.~F.~Brinkman and T.~M.~Rice: \journal{\PRB}{2}{4302}{1970}. 

\bibitem{McMillan} W.~L.~McMillan: \journal{\PR}{138}{A442}{1965}.

\bibitem{Ceperley} 
D.~Ceperley, G.~V.~Chester, K.~H.~Kalos, \journal{\PRB}{16}{3081}{1977}. 

\bibitem{YS1} H.~Yokoyama and H.~Shiba: \journal{\JPSJ}{56}{1490}{1987}.

\bibitem{GA} M.~Gutzwiller, \journal{\PR}{137}{A1726}{1965}. 

\bibitem{MV} W.~Metzner and D.~Vollhardt; \journal{\PRB}{37}{7382}{1988}.

\bibitem{Kaplan} T.~A.~Kaplan, P.~Horsch and P.~Fulde: 
\journal{\PRL}{49}{889}{1982}. 

\bibitem{Fazekas} P.~Fazekas and K.~Penc:
\journal{\IJMP}{B1}{1021}{1988}; P.~Fazekas, 
\journal{Physica Scripta T}{29}{125}{1989}. 

\bibitem{YS} H.~Yokoyama and H.~Shiba:
\journal{\JPSJ}{59}{3669}{1990}.

\bibitem{Y-PTP} H.~Yokoyama: \journal{\PTP}{108}{59}{2002}. 

\bibitem{YTOT} H.~Yokoyama, Y.~Tanaka, M.~Ogata and H.~Tsuchiura: 
\journal{\JPSJ}{73}{1119}{2004}. 

\bibitem{Watanabe} T.~Watanabe, H.~Yokoyama, Y.~Tanaka and J.~Inoue: 
\journal{\JPSJ}{75}{074707}{2006}. 

\bibitem{YOT} H.~Yokoyama, M.~Ogata and Y.~Tanaka: 
\journal{\JPSJ}{75}{114706}{2006}.

\bibitem{Capello} M.~Capello, F.~Becca, S.~Yunoki and S.~Sorella: 
\journal{\PRB}{73}{245116}{2006}. 

\bibitem{YMO} H.~Yokoyama, T.~Miyagawa and M.~Ogata: to appear in 
Physica C (2011), and submitted to \JPSJ. 

\bibitem{Miyagawa} T.~Miyagawa and H.~Yokoyama: to appear in 
Physica C (2011). 


\bibitem{Hubbard} J.~Hubbard: \journal{\PRS}{A267}{237}{1963}.

\bibitem{Kanamori} J.~Kanamori: \journal{\PTP}{30}{275}{1963}.

\bibitem{DMFT} See also, A.~Georges, G.~Kotliar, W.~Krauth and 
M.~J.~Rozenberg: \journal{\RMP}{68}{13}{1996}.

\bibitem{Castellani} C.~Castellani, C.~Di Castro, D.~Feinberg and 
J.~Ranninger: \journal{\PRL}{43}{1957}{1979}. 

\bibitem{Lieb-Wu} E.~H.~Lieb and F.~Y.~Wu: 
\journal{\PRL}{20}{1445}{1968}. 

\bibitem{Harris} For instance, 
A.~B.~Harris and R.~V.~Range: \journal{\PR}{157}{295}{1967}. 

\bibitem{Capello-1D} M.~Capello, F.~Becca, M.~Fabrizio, S.~Sorella 
and E.~Tosatti: \journal{\PRL}{94}{026406}{2005}.

\bibitem{Umrigar} C.~J.~Umrigar, K.~G.~Wilson and J.~W.~Wilkins: 
\journal{\PRL}{60}{1719}{1988}. 

\bibitem{Umrigar-Filippi} C.~J.~Umrigar and C.~Filippi: 
\journal{\PRL}{94}{150201}{2005}; 
S.~Sorella: \journal{\PRB}{71}{241103}{2005}. 

\bibitem{Fletcher} For instance, 
R.~Fletcher: {\it Practical Methods of Optimization} 2nd ed., 
(John Wily, 1987). 

\bibitem{Ibaraki} T.~Ibaraki and M.~Fukushima: 
{\it FORTRAN77 Optimization Programming}, chap.~6 (Iwanami, Tokyo, 
1991), [in Japanese].






\bibitem{HM} C.~S.~Hellberg and E.~J.~Mele: 
\journal{\PRL}{67}{2080}{1991}.

\bibitem{noteWDD} As known from $f_{\rm R}(1)\sim 1$ 
in Fig.~\ref{fig:05Cdd}(b), the repulsive correlation factor 
for the NN sites is useless. 
The reasons why $W_{\rm DD}(1)$ in Fig.~\ref{fig:07PddPdh_multidh}(b) 
is the smallest for $U>U_{\rm c}$ are probably 
(i) a holon(s) occupies the NN sites of the doublon 
[Fig.~\ref{fig:05Cdd}(a)], and 
(ii) the effect of exchange hole is conspicuous for $r=1$. 

\bibitem{notefR} 
$f_{\rm R}(r)$ for $r>4$ (not shown) behaves basically like $f_{\rm R}(4)$, 
but, for $U>U_{\rm c}$, $f_{\rm R}(r)$ slowly decreases to unity 
as $r$ increases. 

\bibitem{note1D} In fact, we have confirmed that A(bind) is insulating 
for any positive value of $U/t$ in the one-dimensional Hubbard model. 



\bibitem{Tahara} For instance, 
D.~Tahara and M.~Imada: 
\journal{\JPSJ}{77}{093703}{2008}, and 
\journal{\JPSJ}{77}{114701}{2008}. 
\end{thebibliography}
\end{document}